\documentclass[usenatbib,usegraphicx]{mn2e}
\usepackage{comment}
\usepackage{amsfonts}
\usepackage{amsmath}
\usepackage{amssymb}
\usepackage{subfigure}

\newcommand{\der}{{\rm d}}
\newcommand{\zs}{z_{\rm s}}
\newcommand{\fzero}{${{\rm d}\widetilde{f_0}(\theta)}/{{\rm d}\theta}$ }
\newcommand{\fone}{$\widetilde{f_1}(\theta)$ }

\newcommand{\eg}{{\it e.g.}~}

\newcommand{\msun}{{\rm M}_{\odot}}

\title[Numerical investigation of lenses with substructures using the perturbative method.]
         {Numerical investigation of lens models with substructures using the perturbative method.}
\author[S.~Peirani et al.]
 {S. Peirani$^{1}$\thanks{E-mail: peirani@iap.fr}, C. Alard$^{1}$,
 C. Pichon$^{1}$, R. Gavazzi$^{1}$ and D. Aubert$^{2}$\\
$^{1}$ Institut d'Astrophysique de Paris, 98 bis Bd Arago, 75014 Paris,
France -\\
 Unit\'e mixte de recherche 7095 CNRS - Universit\'e Pierre et Marie Curie. \\
$^{2}$ Observatoire Astronomique de Strasbourg,  11 Rue de l'Universit\'e, 67000 Strasbourg, France.
}
\begin{document}

\pagerange{\pageref{firstpage}--\pageref{lastpage}} \pubyear{2007}

\maketitle

\label{firstpage}


\begin{abstract}
We present a statistical study of the effects induced by substructures on
 the deflection potential of dark matter halos in the strong
 lensing regime. This investigation is based on the
 pertubative solution around the Einstein radius (Alard 2007) in which all the information on the deflection
 potential is specified by only a pair of  one-dimensional functions on this ring.

Using direct comparison with ray-tracing solutions, we found that the iso-contours of lensed images
 predicted by the pertubative solution is reproduced with a mean error on their radial extension 
 of less than $1\%$ --- in units of the Einstein radius, for reasonable substructure masses. It demonstrates
  the efficiency of the approximation to track possible signatures of substructures. 
 
We have evaluated  these two fields and studied their properties for different lens configurations
 modelled either through massive dark matter halos from a cosmological N-body
 simulation, or via toy models of Monte Carlo distribution
  of substructures embedded in  a triaxial Hernquist potential.

As expected, the angular power spectra of these two fields tend to  have larger values for larger 
harmonic numbers when substructures are accounted for and they can be
 approximated by power-laws, whose values are fitted as a function of the profile and the  distribution 
 of the substructures.
\end{abstract}



\section{Introduction}

 The cold dark matter (CDM) paradigm (Cole et al. 2005
and references therein) has led to a successful explanation of  the 
large-scale structure in the galaxy distribution on scales 0.02 $\leq k \leq$ 0.15h Mpc$^{-1}$.
The CDM power spectrum on these scales derived from 
large redshift surveys
 such as, for instance, the Anglo-Australian 2-degree Field 
Galaxy Redshift Survey (2dFGRS), 
is also consistent with the Lyman-$\alpha$ forest data in the 
redshift range $2\leq z \leq 4$ (Croft et al. 2002;
 Viel et al. 2003; Viel, Haehnelt \& Springel 2004).

In spite of these impressive successes, there are still discrepancies between simulations and 
observations on scales $\leq$ 1 Mpc, extensively discussed in 
the recent literature.
We may mention the sharp central density
cusp predicted by simulations in dark matter halos and confirmed by the rotation 
curves of low surface brightness  galaxies (de Blok et al. 2001) or in bright
spiral galaxies (Palunas \& Williams 2000; Salucci \& Burkert 2000; 
Gentile et al. 2004).
Moreover, deep surveys ($z\geq 1-2$), such as the Las Campanas Infrared Survey,  HST 
Deep Field North and Gemini Deep Deep Survey (GDDS) are revealing an excess of
 massive early-type galaxies undergoing ``top-down'' assembly with high inferred specific
 star formation rates relative to
predictions of the hierarchical scenario (Glazebrook et al. 2004;
Cimatti, Daddi \& Renzini 2006).

One problem that requires closer examination  concerns
 the large number of sub-$L_\ast$ subhalos present in 
simulations but not observed (Kauffmann, White \& Guiderdoni 1993; Moore et al. 1999;
Klypin et al. 1999). This  is the case of our 
Galaxy or M31, although there is mounting evidence for a large number
of very low mass dwarfs (Belokurov et al. 2006).
However, it is still unclear
whether the CDM model needs to be modified to include
 self-interacting (Spergel \& Steinhardt 2000) or warm dark
matter (Bode, Ostriker \& Turok 2001; Col{\'{\i}}n, Avila-Reese \& Valenzuela 2000)
or whether  new physical mechanisms can dispel such discrepancies with the
 observations. For instance, gas cooling can be partly prevented by
 photoionization process which may inhibit star formation in the
 majority of subhalos (Bullock, Kravtsov \& Weinberg 2001).
 
This ``missing satellite problem'' remains an ideal framework
to test cosmological models.
During the past years, different methods have been employed in order to
study the gravitational potential of groups or clusters of galaxies, for instance
through their X-ray lines emission of hot gas in the intra-cluster medium
or through lensing considerations. However, while lensing directly probes the mass
distribution in those objects, the other methods rely more often than
not on strong
hypotheses on the dynamical state of the gas and interactions between baryons and
dark matter. For example, the gas is supposed to be in hydrostatical equilibrium in the
gravitational potential well created by dark matter
halo, while spherical symmetry is assumed.  
In this paper, we study the effects induced by substructures on
the deflection potential of dark matter halos in the strong
lensing regime. The presence of substructures follows from
the capture of small satellites which have not yet been disrupted by
tidal forces and/or suggests that the relaxation of halos is
not totally finished.

Ray-tracing through N-body cosmological simulations suggest that 
substructures should have a significant impact on the formation of giant arcs.
At clusters of galaxies scales, some results indicate that lensing
optical depths can be enhanced
(Bartelmann, Steinmetz \& Weiss 1995; Fedeli et al. 2006; Horesh et
al. 2005; Meneghetti et al. 2007a) whereas some other recent
studies suggest that the impact on arc occurence frequency should only be mild
(Hennawi et al. 2007). On the other hand, the presence of substructures 
changes the properties of strongly lensed images to a point that could lead to
misleading inferred cluster mass properties if not properly accounted for
(Meneghetti et al. 2007b).
Assuming a one-to-one association of cluster subhalos in the mass range
$10^{11} - 10^{12.5}\msun$ and galaxies, Natarajan, De Lucia \& Springel
(2007) used weak lensing
techniques and found the fraction of mass in such subhalos to account for
$10-20\%$ of the total cluster mass, thus in
good agreement with predictions from simulations (Moore et al. 1999).

Likewise, at the scales of galaxies, strong lensing
events have long been involving multiple quasars for which the impossibility
of resolving the size or shape of the lensed images brought the attention toward
flux ratios of conjugate images as a probe of subtructures. Depart from flux
ratios expectations from a
simple elliptically symmetric potential is often interpreted as a signature for
local potential perturbations by substructures
(Brada{\v c} et al. 2002; Dalal \& Kochanek 2002; Brada{\v c} et al. 2004; Kochanek \& Dalal 2004; Amara et al. 2006). It is unclear whether
anomalous flux ratios actually probe ``missing satellites''
(Keeton, Gaudi \& Petters 2003; Mao et al. 2004; Macci{\`o} et al. 2006).
 Due to the small source size
in the case of lensed QSOs, sensitivity to microlensing events due to
stars in the lens galaxy makes the interpretation less obvious.
Astrometric perturbations of multiple quasars have also been considered
(Chen et al. 2007) despite substantial observational limitations.

Presumably the best way out would be to consider extended sources like QSOs
observed in VLBI or lensed galaxies that will be sensitive to a narrower range
of scales for the pertubing potential and thus easier to interprete.
New methods for inverting potential
corrections that needed on top of a smooth distribution were proposed
(Koopmans 2005;  Suyu \& Blandford 2006) but are not guaranteed to converge in all
practical cases and seem to depend on the starting smooth distribution.

One interesting alternative approach is to treat all deviations from
a circularly symmetrical potential as small perturbations (Alard 2007, 2008)
defining the location where multiple extended images will form.
Two perturbative fields, $f_1(\theta)$ and
${{\rm d}f_0(\theta)}/{{\rm d}\theta}$, can then be defined to characterize
deflection potential of lenses as a function of the azimuthal angle $\theta$,
near the Einstein radius. They respectively represent the radial and 
azimuthal derivative of the perturbated potential (see Eq.~(\ref{equ5}) below).
Alard (2008) showed that these two fields have specific properties when one
substructure of mass $\sim 1\%$ of the total mass is positioned near the
critical lines. For instance, the ratio of their {\it angular} power spectra
at harmonic number $n$ is nearly $1$. We will investigate the detailed
properties of these perturbative fields by
considering more realistic lenses such as  dark matter halos
extracted from cosmological simulations. In order to control all the
free parameters (mass fraction, and shapes of subtructures for instance) and
to study their relative impact on arc formation, we will also generate
different families of toy halos.  
 
This paper is organized as follows: in section 2 we present our lensing modelling;
section 3  first sketches the pertubative lens solution
 and applies it  to our simulated lenses for validation against a ray tracing algorithm;
 section 4 presents our main results on the statistics of perturbations,
 while the last section wraps up.

\section{Numerical modelling}
\label{section_modelling}

\subsection{Lens model}

Halos formed in cosmological simulations  tend to
be centrally cuspy ($\rho \sim r^{-1}$) and are generally not spherical,
but have an triaxial shape. The triaxiality of these potential lenses is
expected to increase significantly the number of arcs relative to
spherical models (Oguri, Lee \& Suto 2003 and references therein), and
must be taken into account in numerical models. Thus, 
apart from dark matter halos extracted from cosmological simulations,
we consider in this work typical lenses modelled by of a dark matter halo of total
mass $M = 10^{14} M_{\odot}$  with  
 a generalized Hernquist density profile (Hernquist 1990):

\begin{equation}
\rho(R)=\frac{M}{2\pi}\frac{R_s}{R(R+R_s)^3}\,,
\end{equation} 

\noindent
where $R_s$  is the value of the scale radius, $R$ a triaxial radius
defined by

\begin{equation}
R^2 = \frac{X^2}{a^2} +  \frac{Y^2}{b^2} + \frac{Z^2}{c^2} \,\,\,(c\leq b \leq 1),
\end{equation} 

\noindent
and $c/a$ and $b/a$ the minor:major and intermediate:major axis ratio respectively.
We decided to use an Hernquist
profile
 for practical reasons. However,
for direct comparison with common descriptions of halos from
cosmological simulation in the literature, the Hernquist profile is
related to an NFW profile (Navarro, Frenk \& White, 1996; 1997) with the same dark matter mass within the
virial radius  $r_{200}$\footnote{$r_{200}$ defines the sphere within which the mean 
density is equal to 200 times the critical density.}. Moreover, we also
impose that the two profiles are identical in the inner part ($\leq
R_s$) which can be achieved by using relation (2) between $R_s$ and the
NFW scale radius $r_s$ in
Springel, Di Matteo \& Hernquist (2005).
By convention, we use the concentration parameter
$C_{\rm host}=r_{200}/r_s$  in the following to characterize the density profile of our lenses.
For example, for typical lens at a redshift $z=0.2$, we use 
 $R_s =223$ kpc which corresponds to a NFW profile with $C_{\rm host}=8.0$
 (or equivalently $r_{200}=957$ kpc, $r_s=119$ kpc) and is
consistent with values found in previous cosmological
N-body simulations at the specific redshift and in the framework a the $\Lambda$CDM
 cosmology (Bullock et al. 2001; Dolag, Bartelmann \& Perrotta 2004).

Axis ratios of each lens are 
randomly determined  following Shaw et
al. 2006: $b/a = 0.817 \pm 0.098$,  $c/b = 0.867 \pm 0.067$ and $c/a =
0.707 \pm 0.095$. These values are in good agreement with previous
findings from cosmological simulations (see for instance Warren et al. 1992; Cole \& Lacey
1996, Kasun \& Evrad 2005). 
Finally, it is worth mentioning that each halo is  made of $15
\times 10^6$ particles corresponding to a mass resolution of $6.67
\times 10^6 M_{\odot}$. However, we impose a troncation at a radius of
value $5$ Mpc.

\subsection{Substructures model}

\subsubsection{Mass function}
Numerical N-body simulations show that dark matter halos contain a large
 number of self-bound substructures, which correspond to about 10-20\%
 of their total mass (Moore et al. 1999). 
In the following, 
 the number of substructures N$_{\rm sub}$ in the mass range $m$ -- $m$+$dm$ is
 assumed to obey (Moore et al. 1999; Stoehr et al. 2003)
\begin{equation}
{\rm d} N_{\rm sub} = \frac{A}{m^{1.78}}{\rm d}m\,.
\label{clumpsequation}
\end{equation}
\noindent
The normalization constant $A$ is calculated by requiring the total mass in the clumps
to be $15\%$ of the halo mass and by assuming subhalos masses in the range
$10^{8}$ -- $5\times 10^{12}$ M$_{\odot}$. The minimum number of
particles in the substructures is
about $15$, while the more massive ones have $45,000$ particles.

\subsubsection{Radial distribution}

Substructures are distributed according to the (normalized)
probability distribution
$p(r){\rm d}^3r = (\rho( r)/M_h){\rm d}^ 3r$, where $\rho( r)$ is assumed to
have an Hernquist profile of concentration $C_{\rm sub}$, which yields the probability to
find a clump at a distance $r$ within the volume element ${\rm d}^3r$. 
While the abundance of subtructures in halos of different masses has
recently been extensively quantified in cosmological simulation
 (see for instance Vale \& Ostriker 2004; Kravtsov et
al. 2004; van den Bosh et al. 2007),
their radial distribution is less understood. However, some studies
seem to suggest their radial distribution is significantly less
concentrated than that of the host halo (Ghigna et al. 1998, 2000;
Col{\'{\i}}n et al. 1999; Springel et al. 2001; De Lucia et al. 2004;
Gao et al. 2004; Nagai \& Kravtsov 2005; Macci{\`o} et al. 2006). 
We will use either $C_{\rm sub} = 5.0$ in good agreement with those
past investigations, or $C_{\rm sub}=C_{\rm host}$ for comparison. 

\subsubsection{Density profiles and alignment}

The stripping process caused by tidal forces
seems to reduce the density of a clump at all radii and, in particular, in the central
regions, producing a density profile with a central core (Hayashi et
al. 2003). This process  
 was further confirmed by simulations which found that the inner structure of
subhalos are better described by density profiles shallower than NFW (Stoehr et al. 2003).
However, other simulations seem to indicate that the central regions of 
clumps are well-represented by 
power law density profiles, which remain unmodified 
even after important tidal stripping (Kazantzidis et al. 2004a).
This effect may be enhanced when star formation is taken into account since
dissipation of the gas (from radiative cooling process) and subsequent
star formation lead to a steeper dark matter density profile due to adiabatic contraction.
To test the importance of these differences from the point of view of arcs formation, we
allow for both of these possibilities: we simulate halos with clumps having a
central core $\rho(r) \propto 1/(r_0 + r)^2$, where $r_0$ defines a core
radius, or a central cusp (Hernquist profile) and study how 
our resulting arcs would change from one option to the other. 
It is worth mentioning that each subhalo
concentration parameter is obtained using  relation (13) in Dolag et
al. (2004) within the  $\Lambda$CDM cosmology. However, to avoid spurious
effects due to the lack of resolution, all subhalos  represented by less
than 200 particles will have a concentration parameter value
corresponding to that of  
an halo made of exactly 200 particles (i.e $m=1.33 \times 10^9
M_\odot$).
For core profiles, we follow Hayashi et al. (2003) 
and take a core of size $r_0 \sim r_s$.

Finally, recent cosmological simulations suggest that
subhalos tend to be more spherical than their host (Pereira et al. 2008; Knebe et
al. 2008) and this effect can also be enhanced 
if halos are formed in simulations with gas cooling (Kazantzidis et al. 2004b).
 Moreover, the distribution of the major axes of substructures seems
to be anisotropic, the majority of which pointing towards the center of mass
of the host (Aubert, Pichon \& Colombi 2004; Pereira et al. 2008). Although
shapes and orientations of subhalos provide important constraints on
structure formation and evolution, we have not studied
their relative influence in this work. We reasonably think that modelling
substructures by either triaxial shapes of spherical shapes won't lead
to any significant differences in our results.

\begin{figure}
\rotatebox{0}{\includegraphics[width=\columnwidth]{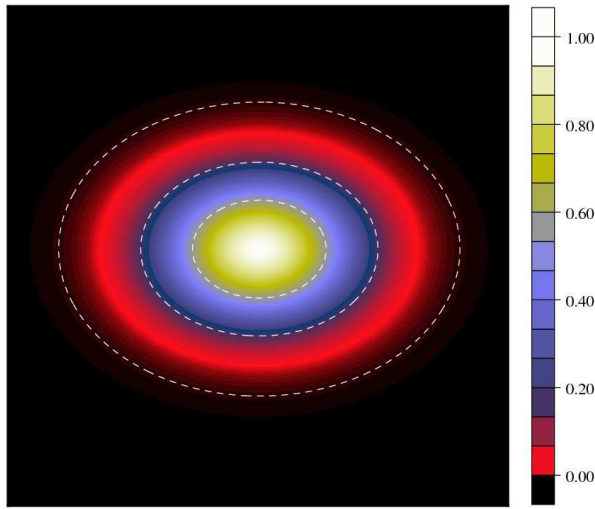}}
\caption{Elliptical luminosity contour of a source with $\eta_0=0.2$. The
 three dashed lines correspond to isophotes defined by $I(R_1) = 0.6\;I_{\rm max}$,
$I(R_2) = 0.2\; I_{\rm max}$ and $I(R_3)=0.01\; I_{\rm max}$. }
 \label{source}
 \end{figure}

\subsection{Lens samples}
\label{lens_samples}

Table 1 summarizes our different samples of lenses. From a statistical
point of view, each sample involves one hundred realizations of halos 
following the above described methodology. Lenses are assumed to be at a
typical redshift $z=0.2$. They share a common total mass of
$10^{14} M_\odot$ and are thus described by an Hernquist profile of concentration
$C=8$. The sample A represents our
reference catalogue in which all halos have no substructure.
For each of them, we introduce a fraction of subtructures ($F_{\rm sub}$) by
removing some background particles so that both the total mass
and the density profile of the initial halo are conserved. These halos are classified
in catalogues B and C according the definition of inner density profile (IP) of clumps.
For example, each halo from samples B1 and B2
 have 15\% of substructures with an
inner profile represented by a cusp (Hernquist profile). Their
radial distribution (RD) within the halo is left as a free parameter. We
consider two possibilities,  $C_{sub}=5$ as suggested by numerical simulations,
and  $C_{sub}=8$, which is the concentration parameter of the whole halo.
Finally, lenses catalogues C1 and C2 have substructures represented by a
core profile with  $C_{sub}=5$ and $C_{sub}=8$ respectively. 

In addition, lensing efficiency depends on the relative distance between lenses and
sources. The efficiency of the lens is scaled by the critical density
\begin{equation}
\Sigma_{crit}=\frac{c^2}{4\pi G}\frac{D_s}{D_d D_{ds}},
\end{equation}
\noindent
where $D_s$, $D_s$ and $D_{ds}$ are the angular diameter distances between
the observer and the source, between the observer and deflecting lens
and between the deflector and the source respectively. When the surface
mass density in the lens exceeds the critical value, multiple imaging occurs.
In order to account this effect which, for a given lensing halo, implies a
different Einstein radius for a different source redshift, we consider the
redshift distribution of sources taken from the COSMOS sample of faint
galaxies detected in the ACS/F814W band (Leauthaud et al. 2007). It is well
represented by the following expression
\begin{equation}\label{eq:zsdistrib}
  \frac{\der n(\zs)}{\der \zs} = \frac{1}{z_0 \Gamma(a)} e^{- \zs/z_0} (\zs/z_0)^{a-1}\;,
\end{equation}
with $z_0=0.345$ and $a=3.89$ (Gavazzi et al. 2007).

\begin{table}
\begin{center}
\begin{tabular}{c|c|c|c}
\hline
 Sample & Sub F& Sub IP & Sub RD\\
\hline
$A$ & 0\% & -  & - \\
\hline
$B1$  & 15\% & cusp & $C_{sub}=5$   \\
\hline
$B2$  & 15\% & cusp & $C_{sub}=8$  \\
\hline
$C1$  & 15\% & core & $C_{sub}=5$  \\
\hline
$C2$  & 15\% & core & $C_{sub}=8$  \\
\hline
\end{tabular}
\end{center}
\caption{Samples of lenses (see details in the text).}
\end{table}


\section{numerical validation of the perturbative solution }\label{section_method}

In this section, we take advantage of the large sample of mock lenses
described in \S~\ref{section_modelling} to assess the validity of the
perturbative method developed in (Alard 2007). After a brief
presentation of
the basic idea in \S~\ref{basics}, we compare the ability of this
simplified procedure to reproduce multiple images lensed by complex
potentials as compared to a direct ray-tracing method
(\S~\ref{rayshoot}) and define the validity range of the perturbative approach.

\subsection{The perturbative approach}\label{basics}
For the sake of coherence, let us sketch the motivation behind 
the perturbative lens
method  (Alard 2007)  used througout this paper. 
The general lens equation, relating the position of an image on the lens plane
to that of the source on the source plane can be written in polar
coordinates as
\begin{equation}
{\bf r_s}=\Big(r-\frac{\partial \phi}{\partial r}\Big){\bf u_r} -
 \Big(\frac{1}{r}\frac{\partial \phi}{\partial \theta}\Big){\bf u_{\theta}}\,,
\label{equ1}
\end{equation}
\noindent
where { $\mathbf{r}_s$} is the source position, and r, {$\mathbf{u}_r$} and {$\mathbf{u}_{\theta}$}
are the radial distance, radial direction and orthoradial direction
respectively. Here $\phi(r,\theta)$ is the projected potential.
Let us now consider a lens with a projected density, $\Sigma(r)$, presenting  circular
symmetry, centered at the origin, and
dense enough to reach critical density at the Einstein radius,
$R_E$. Under these assumptions, the image by the lens of a point source
placed at the origin is a perfect ring, and equation (\ref{equ1}) becomes:
%
\begin{equation}
r-\frac{{\rm d}\phi_0}{{\rm d} r}=0,\,
\label{equ2}
\end{equation}
\noindent
where the potential, $\phi_0$, is a function of r only, and the zero subscript refers to the unperturbed
solution.
The basics ideas of the perturbative approach  is to expand  
equation (\ref{equ2}) by introducing  i) small displacements of the source 
from the origin and ii) non-circular perturbation of the
potential, $\psi$ which can be described by:
\begin{equation}
r_s = \epsilon r_s \,,\quad{\rm and} \quad 
\phi = \phi_0 + \epsilon\psi\,, 
\label{equ3}
\end{equation}
\noindent
where $\epsilon$ is small number: $\epsilon\ll 1$.
To obtain image positions ($r$,$\theta$) by solving equation (\ref{equ1}) directly, may prove to
be  analyically impossible in the general case. It is then easier to find
perturbative solution by inserting equation (\ref{equ3}) into equation
(\ref{equ1}).
For convenience, we re-scale the coordinate system so that the Einstein
radius is equal to unity. The response to the perturbation on $r$ may then  be
written as
\begin{equation}
r=1+\epsilon dr\, ,
\label{equ4}
\end{equation}
which defines
$dr(\theta)$, the azimuthally dependant enveloppe of the relative deflection.
 Using Equation (\ref{equ3}), the Taylor expansion of
$\phi$ is
\begin{equation}
\phi = \phi_0 + \epsilon\psi = \sum_{n=0}^{\infty}[C_n + \epsilon f_n(\theta)](r-1)^n\,,
\label{equ5}
\end{equation}
where:
\begin{equation}
C_n \equiv \frac{1}{n!}\Big[\frac{d ^n\phi_0}{dr^n}  \Big]_{r=1}\,,
\label{equ6}
\quad{\rm
and} \quad
\end{equation}

\begin{equation}
f_n(\theta) \equiv \frac{1}{n!}\Big[\frac{\partial ^n\psi}{\partial r^n}  \Big]_{r=1}\,.
\label{equ7}
\end{equation}
\noindent
Finally, inserting equations (\ref{equ4}) and (\ref{equ5}) into
equation   (\ref{equ1}) leads to:
\begin{equation}
{\bf r_s} = (\kappa_2 {\delta} r - f_1){ \mathbf{u}_r} - \frac{{\rm d}f_0}{{\rm d}\theta}{\bf u_\theta}\,,
\label{equ_contour}
\end{equation}
\label{equ8}
\noindent
with $\kappa_2 = 1-2C_2$.
This equation corresponds to Eq. (8) in Alard (2007).
%

\subsection{ Morphological effects VS astrometric distortions}\label{morpho}

As demonstrated in Alard (2008), the morphology of arcs is very sensitive
 to small perturbators such as substructures in the main halo. The effect
of substructures on morphological features like for instance the size 
of an image is typically much larger than pure astrometric distortions. Indeed, astrometric
distortions are only of the order of the substructure field, which in many
case is smaller than the PSF size of the instrument used, and
 represents then an observational challenge to measure.
Note that a perturbative theory of astrometric distorsions was already
considered by Kochanek  et al. (2001) and Yoo et al. (2005; 2006).
 Note also that the former perturbative approach is
limited to astrometric effects, the relevant theory does not describe image
formation, and thus cannot predict morphological effects.  
When image morphology is considered, effects are an order of
magnitude larger than astrometric effects.
For instance, let us consider the examples displayed in 
Fig.1 and Fig. 2 of Alard (2008).
Whereas a giant arc is obtained from an unperturbed elliptical lens cusp
caustic (Fig. 1), the introduction near the Einstein radius of a substructure
of only 1 \% of the main halo mass breaks the arcs into 3 sub-images (Fig. 2).
The detection of such effects should not require a particularly good resolution,
since the amplitude of the effect is a fraction of the arc size, which is typically
several time the PSF size.

Specifically, let us consider the following example:
we take a sub-critical configuration for an elliptical lens and evaluate 
the modification of the image size due to the perturbation by a substructure field. Let us assume that 
the ellipticity
of the lens is aligned with the axis system and both  source and the
substructure are placed on the X-axis. In this configuration, 
the size of the central image will be perturbed
by the substructure field, and this will be the observable effect.
 The main interest
of this example is the simplicity of the calculations (linear local description of the field)
and the simple description of the effect (reduction of image size).
Let us now
define some useful quantities: the local slope of the unperturbed field $\alpha_0$, the angular 
size of the unperturbed image $\Delta \theta_0$ and the source size  $R_0$.
The substructure parameters are its mass $m_P$ (in unit of the main halo mass) 
and its distance $dr$ to the Einstein ring.
For circular sources, the local slope, image size and source radius are related
by the following relation in the unperturbed case:
\begin{equation}
 \alpha_0 \ \frac{\Delta \theta_0}{2} = R_0\,.
\label{equ_morpho_1}
\end{equation} 
 The slope perturbation introduced
by the substructure $d \alpha$ is (See Alard 2008, Sec. 2):
\begin{equation}
 d \alpha \simeq \frac{m_P}{dr}\,.
\end{equation} 
The substructure field modifies the images size according to:
\begin{equation}
 (\alpha_0 + d \alpha) (\Delta \theta_0 + d \theta) = R_0\,,
\end{equation} 
which gives:
\begin{equation}
|d \theta| = \Delta \theta_0 \frac{d \alpha}{\alpha_0} =  \frac{2  \ m_P \
  R_0}{dr \ \alpha_0^2}\,.
\label{equ_morpho_4}
\end{equation} 
For an elliptical isothermal potential with ellipticity parameter
$\eta$, the field ${{\rm d}f_0(\theta)}/{{\rm d}\theta}$ reads (see Alard 2007, 2008):
\begin{equation}
\frac{d f_0}{d \theta}=\eta \ \sin 2 \theta-\eta \sin \theta\,.
\end{equation} 
Thus, $\alpha_0$ which is the field derivative in 0 is:
$
\alpha_0=\eta\,,
$
leading to:
\begin{equation}
 |d \theta| =  2 \frac{m_P}{dr} \frac{R_0}{\eta^2}\,.
\end{equation} 
The former equation evaluates the angular perturbation of the image size by the substructure. To 
obtain the corresponding perturbation on the image length $dS$, we have to multiply by the Einstein
Radius:
\begin{equation}
 dS \simeq R_E \ |d \alpha| = 2 R_E   \frac{m_P}{dr} \frac{R_0}{\eta^2}\,.
\end{equation} 
Note that astrometric effects are identical to the effects of the substructure on the $f_1$ field
for circular sources. Thus, the astrometric effect $dA$ is of the order (See Alard 2008, Sec. 2):
\begin{equation}
dA \simeq R_E \frac{m_P}{dr}\,.
\end{equation} 
Consequently, the ratio between morphological and astrometric effects here is:
\begin{equation}
\frac{dS}{dA} \simeq \frac{2 R_0}{\eta^2}\,.
\end{equation} 
For arcs, the source size (source diameter= $2 R_0$), and the parameter $\eta$ have the same typical
scale, which gives: $\eta=0.1$, and $R_0=0.05$; then:
\begin{equation}
\frac{dS}{dA} \simeq 10\,.
\end{equation} 
This means that morphological effects are 10 times larger than astrometric effects.
This point is critical, since it really makes the effect of substructure observable.
To illustrate this latter point, let's consider
some numerical values for typical galaxies. For a Milky way
like galaxy, we have a  mass $M \simeq 6 \ 10^{11} M_{\odot}$. The
Einstein radius in arcsec is given by:
\begin{equation}
R_E \simeq 1.8 \sqrt{\frac{M}{10^{12} M_{\odot}}}  = 1.4 \ \ {\rm arcsec}\,,
\end{equation} 
which gives the size of the image perturbation, $dS=R_E \ |d \alpha|$:
\begin{equation}
dS \simeq 2.8  \frac{m_P}{dr}   \frac{R_0}{\eta^2} \ \ {\rm arcsec}\,.
\end{equation} 
For a perturbator with 0.5 \% mass of the galaxy, taking typical
scales, $dr=0.1$, $R_0=0.1$, and $\eta=0.1$,
we obtain:
\begin{equation}
dS \simeq 1.4  \ \ {\rm arcsec}\,.
\end{equation} 
Such effects should be within the  reach of spatial instruments such as DUNE or
SNAP. On the contrary,  astrometric
effects are typically 10 times smaller $dA \simeq 0.14 \,\,{\rm arcsec}$ and
should be then much more difficult to detect.
Consequently morphological effects are definitely our best hopes to detect substructures.

Clearly,  other configurations will also
allow the measurement of morphological effect of substructures, though in
general the calculations will 
be a little bit more complicated, but the effects will more or less be
of the same order, since they are 
related to the amplitude of the perturbation of the ${{\rm d}f_0(\theta)}/{{\rm d}\theta}$ field by the
substructure.


\subsection{Reconstruction of images}\label{rayshoot}

\begin{figure*}
\rotatebox{0}{\includegraphics[width=18cm]{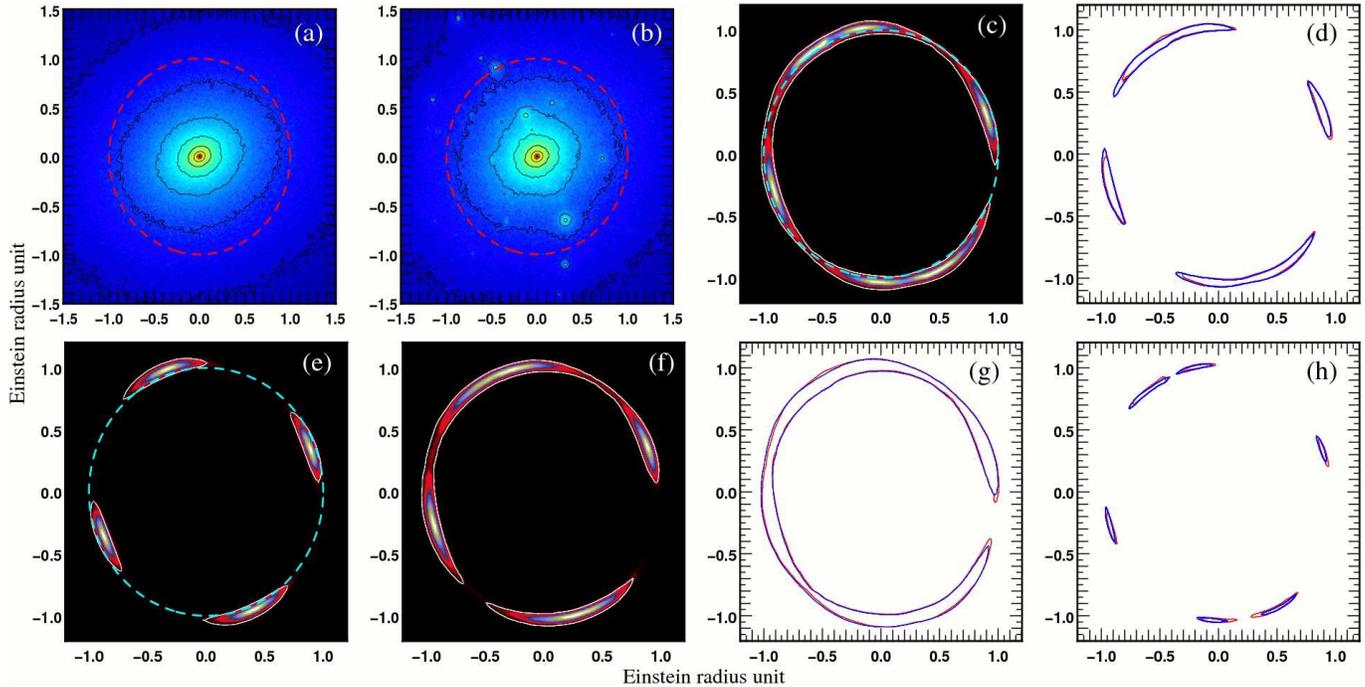}}
\caption{Projected density maps of lenses modelled by toy halos and
 their associated arcs reconstructions when considering an elliptical
 source contour with a Gaussian luminosity.
 Panels (a) and (e) show respectively the projected density map of the lens $L_0$ and the
 resulting image obtained from ray-tracing  when the source is placed at
 the origin.  Both red and light blue dashed lines represent
 the Einstein radius. The solid white line is the predicted arc
 reconstruction for isophotes  $0.01 I_{\rm max}$.
 Similar plots are shown in panels (b) and (f) but for
 $L_3$. Columns 3 and 4 show arc reconstructions for the lens
 $L_2$. The solution from ray-tracing is given in the panel (c)
 while the other last panels present a direct comparison between
 the isophotal contour (red) and the arc
 reconstruction (blue) for $0.01 I_{\rm max}$ (g),
 $0.2 I_{\rm max}$ (d) and  $0.6 I_{\rm max}$ (h).  
  }
 \label{fig_s2}
 \end{figure*}

One interesting feature of the perturbative method is to provide a
framework for the reconstructions of images.
By first defining an elliptical source centered on position $(x_0,y_0)$, with
a characteristic size $R_0$, ellipticity $\sqrt{2 \eta_0}$, and inclination of the main
axis $\theta_0$ such that,

\begin{eqnarray}\label{eq:source}
  R_0^2&=&(1-\eta_0 \cos \theta_0)(x_s-x_0)^2  \cr \cr
   &+& 2 \eta_0 \sin 2 \theta_0 (x_s-x_0)(y_s-y_0)   \cr \cr
   &+& (1+\eta_0 \cos \theta_0) (y_s-y_0)^2  
\end{eqnarray}
%
one can express the equations of the image contours using Eq.~(\ref{equ8}): (see Alard (2007) for more
details)
%
\begin{eqnarray}\label{equ_contour}
dr^{\pm}&=&\!\!\frac{1}{\kappa_2}\left[\widetilde{f_1} + \sin 2 \widetilde \theta \frac{\eta_0 }{S}\frac{d\widetilde{f_0}}{{\rm d}\theta}     \pm
 \frac{\sqrt{R_0^2S - (1-\eta_0^2)({d\widetilde{f_0}}/{d\theta})^2}}{S}  \right]  \cr \cr
S\!\!&\equiv&\!\!1-\eta_0 \cos 2 \widetilde \theta\,, \cr \cr
\widetilde \theta\!\!&\equiv&\!\! \theta-\theta_0
\end{eqnarray}
This equation corresponds to Eq. (15) in Alard (2007). 
The functional $f_i$ is defined to take into account the effect of the  translation of
the source  by the vector ${\bf r_0}=(x_0,y_0)$.
\begin{equation}
\widetilde{f_i} = f_i + x_0 \cos\theta + y_0 \sin\theta, \,\,\,\, {\rm for}\,\,\,\, i=0,1.
\label{translation}
\end{equation}
As emphazised in Alard (2007), the image contours are only governed by 
the two fields,  \fone and \fzero,
which contain all the information on the deflection potential at this order in the perturbation.
For instance, the two first terms in the bracket of
Eq. (\ref{equ_contour}) give informations on the mean position of the
two contour lines while the last term provides  informations on the image's
width along the radial direction as well as a  condition for image formation. 
Therefore, the  characterization of these two fields represents a simple and
efficient way to track possible signatures of the deflection potential
induced by  substructures, so long as the perturbative framework holds, as 
we will illustrate below.
Alard (2007) already implemented the method 
 with a lens described by a NFW
profile yielding an analytical solution for the projected potential profile.
In this section, we  illustrate and validate the method while considering more complicated
and realistic situations. In particular, we use lenses  either from cosmological
simulations or from toy models presented in section \ref{section_modelling}.

For direct comparison between arc reconstructions predicted by the
perturbative method and theoritical ones, we use a ray-tracing method. 
Part of our investigations indeed makes use of  the Smooth Particle Lensing
 technique (SPL),  described in details in Aubert, Amara \& Metcalf
 (2007) and summarized in this section. SPL has been developed to
 compute  the gravitational lensing signal produced by an arbitrary
 distribution of particles, such as the ones provided by numerical
 simulations. It describes particles as individual light deflectors
 where their surface density is arbitrarily chosen to be 2D Gaussian.
 This choice makes it possible to compute the analytical corresponding
 deflection potential, given by:
\begin{equation}
\phi(r)=\frac{m_p}{4\pi\Sigma_c}(\log(\frac{r^4}{4\sigma^4})-2\mathrm{Ei}(-\frac{r^2}{2\sigma^2})),
\end{equation}
where $\mathrm{Ei}(x)=-\int^{\infty}_{x} \exp(-x)/x \mathrm{d}x$,
 $m_{p}$ is the mass of the particle, $\sigma$ its extent and
 $\Sigma_{c}$ the critical density. From the deflection potential,
 expressions for the deflection angles $\bf \alpha$, the shear
 components $\bf \gamma$ and the convergence $\kappa$ can be easily
 recovered (see Aubert, Amara \& Metcalf (2007) for more details). Knowing the
 lensing properties of a single particle, one can recover the full
 signal at a given ray's position on the sky by summing the
 contributions of all the individual deflectors:
\begin{eqnarray}
\phi(\mathbf{r})&=&\sum_i \phi_i(\mathbf{r}) \,,\quad
{\vec{\alpha}}(\mathbf{r})=\sum_i  {\vec{\alpha}}_i(\mathbf{r})\,,\\
\kappa(\mathbf{r})&=&\sum_i  \kappa_i(\mathbf{r}) \,,\quad
{\vec{\gamma}}(\mathbf{r})=\sum_i  {\vec{\gamma}}_i(\mathbf{r}),
\end{eqnarray}
where, {\sl e.g.} $\vec\gamma_{i}(\mathbf{r})$ is the contribution of the i-th
 particle to the shear at ray's position $\mathbf{r}$. This summations are
 performed efficiently by means of 2D-Tree based algorithm, in the
 spirit of N-body calculations. The tree calculations are restricted to
 monopolar approximations where an opening angle of $0.5-0.7$ is found to
 give results accurate at the percent level on analytical
 models. Finally, Aubert, Amara \& Metcalf (2007) found that an adaptative
 resolution (i.e. an adaptative extent $\sigma$ for the particles)
 provides a significant improvement in the calculations in terms of
 accuracy. For this reason, the smoothing $\sigma$ depends on the rays
 location:  particles shrink in high density regions in order to
 increase the resolution while they expand in low-density regions,
 smoothing the signal in undersampled areas.  
For all simulations, we use $1024 \times 1024$ rays within a square of
size $2\times 2 R_E$, an opening angle of 0.7 and $N_\sigma=256$
(where $N_\sigma$ is the number of particles over which the smoothing is
applied).

\subsubsection{Lenses from the toy model}
\label{toy_model}

We present in this section characteristic examples of
arc reconstruction.
Three lenses $L_0$, $L_1$ and $L_2$ belonging to samples A, B2 and C2
respectively are considered. They have a common mass, density profile, axis ratios and
random orientation in 3D space. They only differ via the presence or not
of substructures as well as via the inner density profile of substructures: 
$L_0$ has no substructure whereas $L_1$
has substructures with a central cusp while a core describe the inner
density profiles of substructures in $L_2$. 
In the present case, the source is at a redshift $z\sim 2.9$,  has an
elliptical contour with $\eta_0 =0.2$ and a radius $R_0\sim0.05R_E$, characterizing
a Gaussian luminosity profile (see figure \ref{source}).

For the arc reconstructions presented below, we shall consider 3 different
radii $R_1$, $R_2$ and $R_3$ corresponding to 3 specific isophotal contours
defined by $I(R_1) = 0.6\; I_{\rm max}$, $I(R_2) = 0.2\; I_{\rm max}$ and
$I(R_3)=0.01\; I_{\rm max}$ (see figure \ref{source}).

\begin{figure}
\rotatebox{0}{\includegraphics[width=8.5cm]{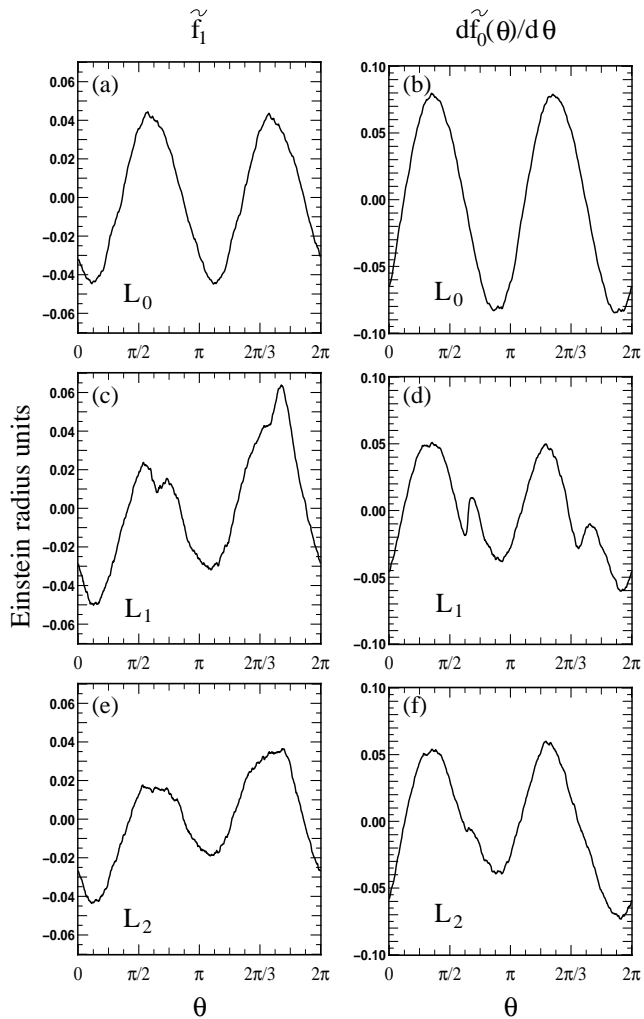}}
\caption{Variations of the fields $f_1$  and 
${{\rm d}f_0(\theta)}/{{\rm d}\theta}$  as a function of
$\theta$ for the lens $L_0$ (panels a and b), lens $L_1$
(panels c and d)  and lens $L_2$  (panels e and f).}
 \label{fig_champs}
 \end{figure}

In Figure (\ref{fig_s2}), we show the projected mass density of lenses $L_0$
and $L_1$  near
the Einstein radius, the image's solution obtained from ray-tracing and the
contours predicted by the perturbative method when the source is placed at the origin.  
When no substructure is considered, both projected density and potential are  
nearly elliptical. As expected, we obtain  four distinct arcs in a cross configuration. The
predicted arcs reconstruction are in good agreement with the numerical solution
obtained via the  ray-tracing algorithm.  From a theoretical point of
view, it is easy to show that the functions \fone and \fzero
are proportional to $\propto \cos(2\theta + \psi)$ and $\propto
\sin(2\theta + \psi)$ respectively.     
These functionnal form are recovered in our experiment and shown in Figure (\ref{fig_champs}).

When substructures are present, the shape of images is significantly
altered. First, we notice that the positions  of substructure tend to break
the ellipticity of the halo center. Thus, it is not surprising that the
shape of the image is approaching a ring in that case. 
Moreover, it is interesting to see that the position of one substructure (at the top left
of the figure) is exactly at the Einstein radius. This produces an
alteration of the luminosity, while the  effects
is more violent when substructures present a cups
profile. This effect can be clearly seen when comparing the perturbative fields
\fone and \fzero relative to the lens $L_1$
in Figure (\ref{fig_champs}). For instance, we can see two clear bumps in the
evolution of \fzero. The second one ($\theta >2\pi /3$) 
is produced by substructures in the lower right part and
induced an alteration of the luminosity again.

To estimate the systematic error between the theoretical contours
provided by the ray-tracing and
those predicted from equation (\ref{equ_contour}), we use a simple
procedure with a low computational cost. First,
each predicted contour is divided into a sample of N points.
Each of them is defined by  polar coordinates ($r_i$, $\theta_i$)
which  coincide with a luminosity value of the image
corresponding to an unique radius $R_i$ in the source frame. By using
relation (\ref{equ_contour}), we then compute
$1+dr(R_i)$ which gives the image contour radius of the isophotal contour
$I(R_i)$ of the source in Einstein radius unit. By defining $1+dr_i$ the
radial distance of point i, the mean error ${\rm err}$ (in Einstein radius unit) is then computed by
${\rm err} = \sum_i^N|dr(R_i) - dr_i|/{N}$.  

For illustration, we have estimated the mean error reached   for the lenses
$L_1$ and $L_2$ using three luminosity contours source (see fig. \ref{source}).  
For $L_1$,  the mean errors are respectively 0.67\%, 0.71\%, 0.95\% of the  Einstein radius  
for isophotes $0.6$, $0.2$ and $0.01 I_{\rm max}$ respectively,  while we obtain
 0.74\%, 0.86\% and 1.04\% $R_E$  for lens $L_2$ and for same
 luminosities. 
We have also studied how the mean error evolves for random positions of the
source inside an area limited by the caustic lines.  To do that, we have
  used the lens $L_1$ and have studied 100 realizations with different impact parameters.
We found ${\rm err} = (1.01 \pm 0.12) \% R_E$  for isophotes
equal to $0.01 I_{\rm max}$.

\subsubsection{Lenses from cosmological simulation halos}
\begin{figure}
\rotatebox{0}{\includegraphics[width=\columnwidth]{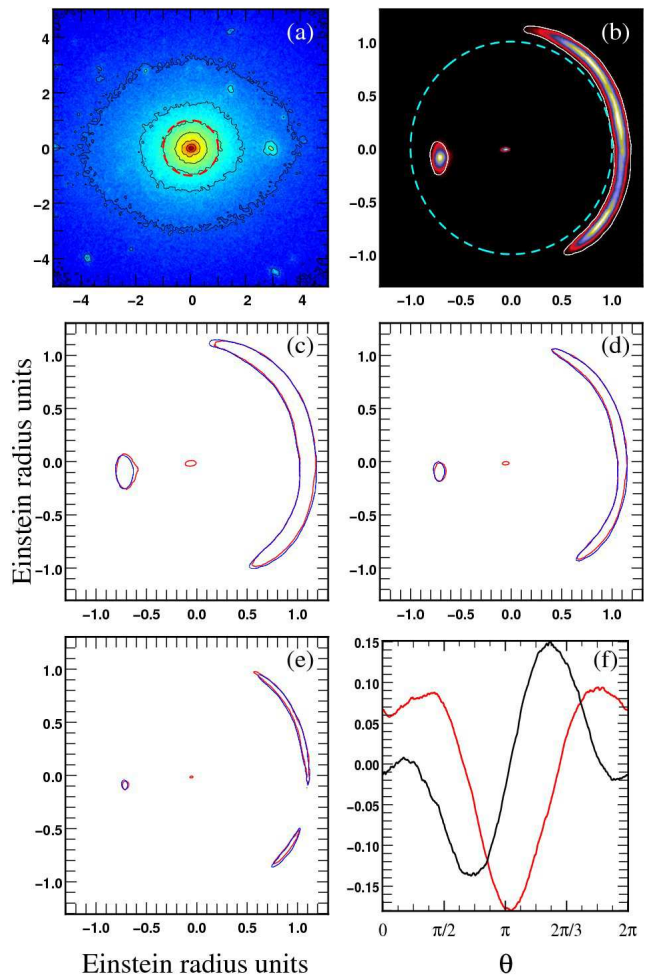}}
\caption{Projected density map (panel a) of a lens modelled by a dark matter halo
 extracted from the N-body simulation and the  associated arc
 reconstructions. The solution from the  ray-tracing is plotted in the 
 panel (b) with the Einstein radius (blue line) and
 the predicted arc reconstruction for isophote $0.01 I_{\rm max}$. The second
 line compares the isophote contours $0.01 I_{\rm max}$ (panel c) and
 $0.2 I_{\rm max}$ (panel d) represented by the blue lines with the
 predicted contour (red lines). The same results are presented in the
 panel e for isophote $0.6 I_{\rm max}$. Finally, variations of 
\fone (red line) and \fzero ({\sl black line})
 are plotted in the panel (f)).
}
 \label{fig3}
 \end{figure}

In this section, lenses are modelled by 
dark matter halos extracted from a cosmological simulation of the 
{\tt Projet HORIZON}\footnote{http://www.projet-horizon.fr/}.
The simulation  was   run  with {\tt Gadget-2}  (Springel 2005) for  a   $\Lambda$CDM   universe   with
$\Omega_M=0.3$,  $\Omega_{\Lambda}=0.7$, $\Omega_B=0.045$,  $H_0=70$ km/s/Mpc,
$\sigma_8=0.92$ in  a periodic box  of 20 $h^{-1}$Mpc. We use $512^3$
particles corresponding to a 
mass resolution of $m_{\rm part.}  \simeq  4 \times 10^6 M_{\odot}$
and a spatial resolution of 2 kpc ({\it physical}).
Initial conditions has been generated from the MPgrafic code (Prunet et
al. 2008), a
parallel (MPI) version of {\tt Grafic } (Bertschinger 2001). 
In this simulation, we selected two regions. In the first one, the lens is a typical
halo of total mass $ 2.6 \times 10^{13} M_\odot$ at
redshift $z=0.5$. The source is at $z=1.2$, assumed to be elliptical
($\eta_0=0.2$) with $R_0=0.05 R_E$ and placed near a caustic in order to obtain a giant arc.  
Fig. (\ref{fig3}) shows the projected density of the lens near the
Einstein radius; both the ray-tracing solution and
the predicted contours by the perturbative method are shown.
Here again, the three different contours are well reconstructed since
the error are 0.76\%, 0.83\% and 0.91\% $R_E$ for isophotes $0.6$, $0.2$ and
$0.01 I_{\rm max}$ respectively. For illustration, the angular variation of the
perturbative fields are also representated in the figure (\ref{fig3}).

\begin{figure}
\rotatebox{0}{\includegraphics[width=\columnwidth]{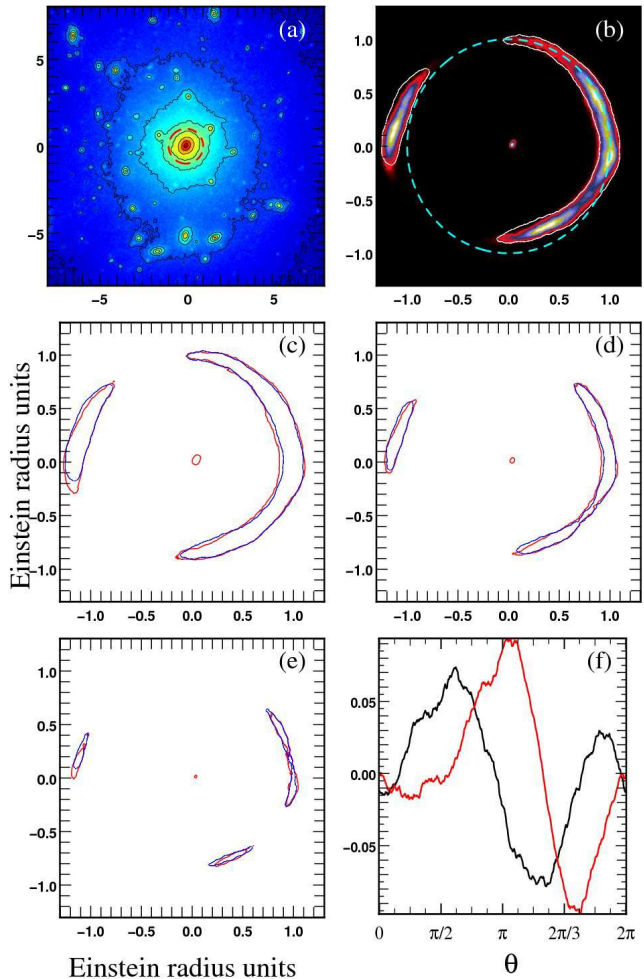}}
\caption{
same as Figure \ref{fig3} for  another dark matter halo extracted from the N-body simulation.
 }
 \label{fig4}
 \end{figure}

The second example is a lens modelled by another halo from the same N-body
simulation. Its total mass is $6.6 \times 10^{13} M_\odot$ at  $z=0.5$.
This is an extreme case since a significant number of substructures are
still falling toward the center of the host halo which suggests that the dynamical
relaxation is still operating. This strongly affects the potential and
the perturbative fields (see figure \ref{fig4}). 
However, mean errors remain small of the order of
1.11\%, 1.20\% and 1.26\% $R_E$ for isophotes $0.6$, $0.2$ and
$0.01 I_{\rm max}$ respectively, which proves the accuracy of the method to
deal with more complex systems. 
We may reasonably  think that lenses in our different samples at $z=0.2$
tend to be  more relaxed that the present configuration, and, consequently, the
mean errors in our arc reconstruction should be less pronounced.

\section{Fourier series expansion}\label{section_statistics}

\subsection{Motivation}
Given Eq. (\ref{equ_contour}), it is straightforward,  for a given lensed image and an assumed
underlying spherical lense and elliptical source, to invert it for \fone and \fzero as 
\begin{eqnarray} \label{eq_invert_0}
\frac{{\rm d}{\tilde f}_0}{{\rm d}\theta}&=& \!\!\!\frac{ \pm S}{2 \sqrt{1-\eta_0^2}} \sqrt{4 R_0^2/S-\kappa_2^2[ dr^+(\theta)-dr^-(\theta)]^2}\\ \label{eq_invert_1}
{\tilde f}_1(\theta)&=&\!\!\!\frac{1}{2}\left( \kappa_2[dr^+(\theta)+dr^- (\theta)]-2  \frac{{\rm d}{\tilde f}_0}{{\rm d}\theta} \frac{\eta_0 \sin 2 \widetilde \theta}{S} \right) \,,
\end{eqnarray} where $S$, $\widetilde \theta$ and $\kappa_2$ are given by  Eq.~(\ref{equ_contour}).
%
This inversion formula depends explicitly on the source parameters,
 $(\eta_0, \theta_0)$ which are unknown.
However using Eqs (~\ref{eq_invert_0}, ~\ref{eq_invert_1}) it
is possible to compute the two functions \fone and \fzero for
each couple of parameters $(\eta_0, \theta_0)$. The proper solution
corresponding to the true parameter
$(\eta_0, \theta_0)$ has minimal properties. Consider for instance
a circular solution ($\eta_0=0$), if the
inversion formula is used with $\eta_0 \neq 0$ additional Fourier
terms with order $n>2$ will appear in the inversion
formulae. Thus, it is clear that minimizing the power in higher
order Fourier modes is a criteria that will allow
to select the best solution when exploring the plane $(\eta_0,
\theta_0)$.
This criteria has also a very interesting
property, considering that power at order $n>2$ usually reveal
the presence of substructures (see Table 3 for instance),
the solution with minimum power at higher order is also the one that
puts the more robust constraint on the presence of substructure. Thus
the elliptical inversion can be performed by exploring the plane
$(\eta_0, \theta_0)$ in a given
parameter range, computing the corresponding fields \fone and \fzero
and their Fourier expansion, and selecting
the solution with minimum power at $n>2$.
%
For non elliptical sources, one can use the general inversion method presented in Alard (2008). 
This inversion method remaps the images to the source plane using 
local fields models (basically the scale of the images).
The solution is selected by requiring maximum similaritiy of the images in the source
plane. Image similarity is evaluated by comparing the image moments up to order $N$. 
Provided the number of image moments equations exceed the number of model
parameters, the system is closed and has a definite solution.  Note
that the local models may be replaced
with general Fourier expansion in the interval $0<\theta<\pi$, but in
this case, the additional constraint
that no image are formed in dark areas must be implemented (Diego et al. 2005). 
In the perturbative approach this requirement can be reduced to \fzero$>R_C$
in dark areas, where $R_C$ is the radius of the smallest circular contour
that contains the source.
We may therefore assume for now that observational data may be inverted, and
that an observationnal survey of arcs should provide us with a statistical
distribution of the perturbatives fields. 
Hence we may use our different samples of halos
presented in paragraph \ref{lens_samples} in order to measure the
 relative influence on arc formation of the different free parameters such as
the inner profile of substructures or their radial distribution within the host
halo. To conduct this general analysis the fields will be represented by Fourier models,
due to the direct correspondance between Fourier models of the fields and the multipolar 
expansion of the potential at $r=1$ (Alard 2008).
%
%
%
\subsection{Results}
The angular functions \fone and \fzero can be characterized by their Fourier
expansion:
\begin{eqnarray}
\frac{d\widetilde{f_0}(\theta)}{{\rm d}\theta} &=& \sum_n \langle a^0_n \rangle \cos \left(n\theta+\phi^0_n\right) \,, \\
\widetilde{f_1}(\theta)&=& \sum_n \langle a^1_n \rangle  \cos \left(n\theta+\phi^1_n\right) \,,\\
P_i(n)  &=& \langle {(a^i_n)}^2 \rangle, \,\,\,\, {\rm where} \,\,\,\, i=0,1,
\label{}
\label{expansion_multi}
\end{eqnarray}
where $P_i(n)$, $i=1,2$ correspond to associated power spectra.
We have derived the multipole expansion of \fone and \fzero for each
halo of the different catalogues and we focus in the following on the
mean amplitudes $\langle a^{0}_n \rangle$ and $\langle a^{1}_n \rangle$ obtained.

Tables (\ref{tab:f1tab}) and (\ref{tab:f0tab}) respectively summarise
the seven first orders of the power spectrum of \fone and
\fzero for the 3 lenses $L_0$, $L_1$ and $L_2$.

\begin{table}
\begin{center}
\begin{tabular}{c|c|c|c|c|c|c|c}
\hline
 Lens & 1& 2 & 3 & 4& 5 & 6& 7\\
\hline
$L_0$ & 0.07 & 4.21 & 0.02  &0.20  &0.04  & 0.07  & 0.03\\
\hline
$L_1$  &1.62 & 3.80  & 0.42 &0.18  &0.29   & 0.20 &0.33  \\
\hline
$L_2$  &  1.38  & 2.86 & 0.18 & 0.20  & 0.10& 0.11 &0.11 \\
\hline
\end{tabular} 
\end{center}
\caption{Power spectra of \fone shown in the first column of Figure~\ref{fig_champs} \label{tab:f1tab}.}
\end{table}
\smallskip
 
\begin{table}
\begin{center}
\begin{tabular}{c|c|c|c|c|c|c|c}
\hline
 Lens & 1& 2 & 3 & 4& 5 & 6 & 7\\
\hline
$L_0$ &0.08  &8.17  & 0.04  & 0.39 & 0.02 & 0.08  &0.03 \\
\hline
$L_1$  &1.14 & 4.12  &0.32  & 1.50 &  0.28  & 0.59 & 0.24\\
\hline
$L_2$  &1.54 & 5.36  & 0.20 &  0.74 & 0.07  &0.14 &0.18 \\
\hline
\end{tabular}
\caption{Power spectra of \fzero  shown in the second column of Figure~\ref{fig_champs} \label{tab:f0tab}.}
\end{center}
\end{table}

When substructures are absent, both harmonic power spectra of \fone
 and \fzero are dominated by the second order mode, 
which is characteristic of a projected elliptical potential. The situation is totally different when
substructures are taken into account. First, we notice that first mode ($n=1$) 
increase for lenses $L_1$ and $L_2$. This is due to
the fact that we kept the same definition
of the mass center between the three lenses. The random position of subtructures
 generates a non zero impact parameter which affect the first  order mode
according the relation (\ref{translation}).
Moreover, since substructures tend to break the
ellipticity of the halo center in the present case, one expects 
 that the second mode decreases. However, the most interesting feature is that modes corresponding to 
  $n\geq3$
increase when substructures are present.

\subsection{Practical limitations}\label{noise}

\begin{figure}
\rotatebox{0}{\includegraphics[width=8cm]{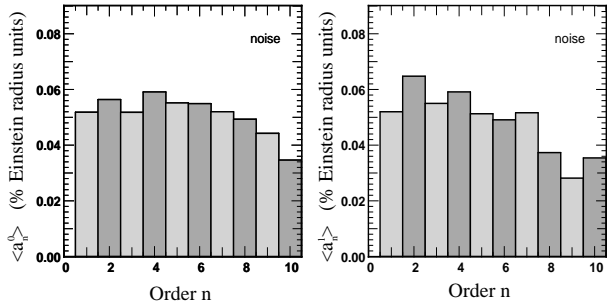}}
\caption{Variations of the mean amplitudes,  $\langle a^0_n \rangle $
 ({\sl left panel}), and 
  $\langle a^1_n \rangle $ ({\sl right panel}), as a function of the harmonic  order, $n$, derived from 100 lenses modelled
 by an isothermal profile.}
 \label{noise_fig}
 \end{figure}

 A fraction of the error is produced by the ray-tracing
simulation as well as the limitation of the considered resolution. 
Let us therefore consider a
toy halo with an isothermal profile:
\begin{equation} 
\rho = \frac{\rho_0}{r^2},
\label{isothermal}
\end{equation}
where $\rho_0$ is evaluated so that the mass enclosed inside a radius
$r=957$ kpc is $M=10^{14} M_\odot$. An isothermal profile is appropriate
to estimate systematic error since it leads to exact
solution with the perturbative method. 
Here again, we have evaluated the quantity ${\rm err}$ by considering 100 realizations
with different impact parameters. Sources have circular contour with
$R_0 \sim 0.05 R_E$ and 15 millions particles have been used.
For isophotes $0.6 I_{\rm max}$, which is supposed to have the
higher error values, we have obtained
${\rm err} = (0.30 \pm 0.03)$ \% $R_E$. Thus, in the following, we will consider
that both ray-tracing method and the resolution  limitation  induce to a mean
error of 0.3 \% $R_E$ in contours reconstructions.
Moreover, as we shall see in section \ref{section_statistics}, \fone and \fzero can be
characterized by their multipole expansion and their  associated power spectrum (see equations
\ref{expansion_multi}). In Figure (\ref{noise_fig}), we plot
the amplitudes $\langle a^0_n \rangle $ and $\langle a^1_n \rangle $   as a function of n derived from
the present experiments. These values for the amplitude ($\sigma \sim 0.06 \% R_E$) correspond to a noise 
that we have to take into account below. For
this reason, we put a confident limit to $\sim 2.0 \sigma$.    

In Aubert et al. 2007, the influence of smoothing $N_\sigma$,
number of particles  $N_{part}$
and opening angles $\theta$ have been extensively investigated on softened
isothermal spheres. The number of particles and the models used here
are similar and the parameters used in the current study can be
considered as the most appropriate considering these previous tests
($N_\sigma=256$, $\theta=0.7$, $N_{part}=15\times10^9$). For instance a re-analysis of the Aubert
et al. tests case present an average error in the deflection angle of
0.3-0.4 \% at the Einstein radius. Considering that the profiles and
the number of particles are similar, the current error estimation is
in good agreement. 

In the same studies, the (inverse) magnification
reconstruction was also previously tested and the critical lines
fluctuates around their theoretical location because of Poisson noise
and ray-shooting artefacts. By increasing the number of particles and by
means of adaptive smoothing, these errors can be limited. Again, the
re-analysis of these tests cases allows an estimation of the error on
magnification $\mu$  of $\delta \mu/\mu \sim 0.02 \mu$ close to the
Einstein's radius. The inverse magnification is obtained from the joint
calculation of the convergence and the shear through the same
ray-shooting technique. Hence, the error estimation does not rely on
some propagation procedure but on the effective calculation and thereby
includes Poisson sampling effects and ray-shoot errors.

To finish, it's interesting to determine how the error of
 0.3 \% $R_E$ in arc reconstruction is reflected in error in the image length.
Let's consider again the configuration studied in Sect. \ref{morpho} and
taking  equations (\ref{equ_morpho_1}) and (\ref{equ_morpho_4}), one obtains:
\begin{equation}
|d \theta| = 2 \Delta \theta_0 \frac{d \alpha}{\alpha_0} = 2 \frac{d
  \alpha \ R_0}{\alpha_0^2}  
\end{equation} 
Taking $d \alpha=0.3/100$ with the same typical value, $ R_0=\eta
= 0.05$, and $\alpha_0=\eta=0.1$ we have:
\begin{equation}
 |d \theta| = \frac{3}{100}
\end{equation} 
As in the previous calculation, the actual size of the perturbation
due to the error $d_E$ is obtained
by direct multiplication with the Einstein radius. We take the same
numerical value, $R_E=1.4 \ \ {\rm arcsec}$,
thus:
\begin{equation}
d_E=0.04 \ \ {\rm arcsec}
\end{equation} 
Given the amplitude of morphological effects, this noise is not a source of concern, but
 had we considered astrometric effects, this noise would become
be a real problem.

\subsection{Statistics}\label{statistics}
\begin{figure*}
\rotatebox{0}{\includegraphics[width=17.5cm]{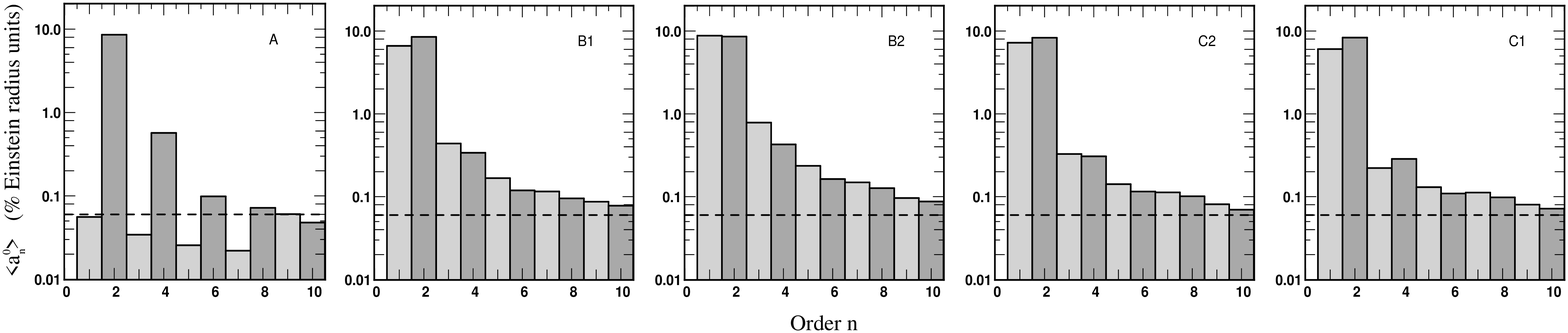}}
\caption{Variation of mean amplitudes $\langle a^0_n \rangle $ derived from multipole
 expansions of \fzero and for each lenses catalogue. The dashed line
 represent limits at $1\sigma$.}
\label{figf0}
\end{figure*}

\begin{figure*}
\rotatebox{0}{\includegraphics[width=17.5cm]{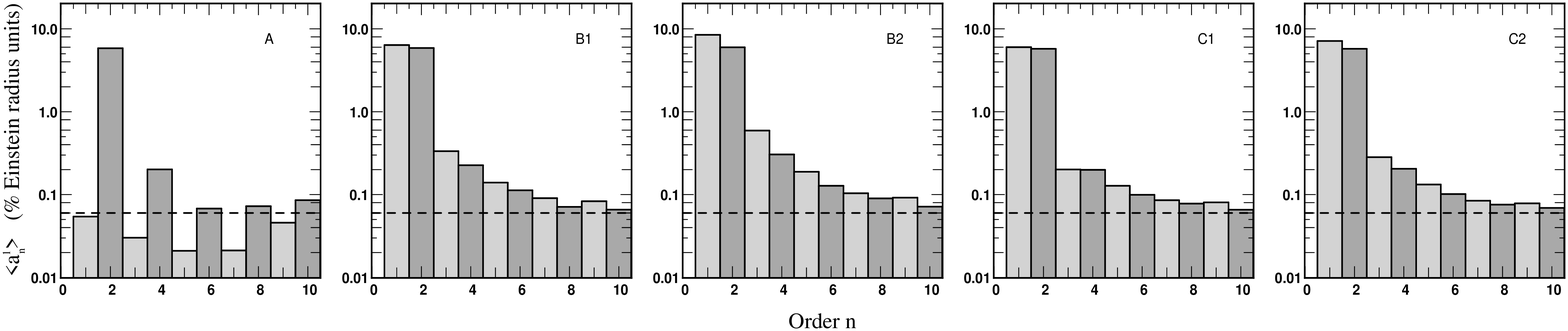}}
\caption{Variation of mean amplitudes $\langle a^1_n \rangle $ derived from multipole
 expansions of \fone and for each lenses catalogue. The dashed lines
 represent limits at $1\sigma$.}
\label{figf1}
\end{figure*}

Figures (\ref{figf0}) and  (\ref{figf1}) show respectively the
variations of the mean
amplitudes $\langle a^0_n \rangle $ and $\langle a^1_n \rangle$ as a function of $n$ 
derived from our different samples of lenses.
As mentioned above, we put a confident limit  ($\sim
 2.0\sigma = 0.12 \% R_E$)
 and we exclude in our calculation all amplitudes below.
 First, when
substructures are disregarded, power spectra are dominated by  the second harmonic
which just reflects the fact  that our simulated lenses have a mean  ellipticity.
However, values of  the fourth orders appear to be 
 not negligible. This is probably due to the 2D projection 
which can lead to boxy projected densities. On the other hand, odd
orders are negligible, as expected. 
When substructures are present, we can  notice that  the amplitudes
of  first harmonic, $\langle a^i_1 \rangle$, have rather high values. As emphasized before,  
this is simply due  to
the fact the position of the lenses center is assumed to be the same for all objects.
In fact,  subtructures 
 tend to modify the position of the mass center or, equivalently, tend
 to generate a non zero impact parameter which affect first harmonic coeficient  
according to the relation (\ref{translation}).
However, the most interesting and important result is the presence of a tail in the power
spectra which clearly suggests that  the amplitude of high order  harmonics ($n\geq 3$)
are not negligible anymore. Note that these effects are clearer when the density profile of
substructures is modelled by a cusp as was the case for sample $B2$. 
Note that the power spectra of high orders harmonic ($n \geq 3$) can be fitted by power-laws:
\begin{equation}
P_0(n\geq 3) = k_0 n^{\alpha_0}  \,, \quad
P_1(n\geq 3) = k_1 n^{\alpha_1}  \,,
\label{eq:fit}
\end{equation}
\noindent
where the relevant fitting
parameters are shown in the Table~\ref{tab:Q}.

\begin{figure*}
\rotatebox{0}{\includegraphics[width=18cm]{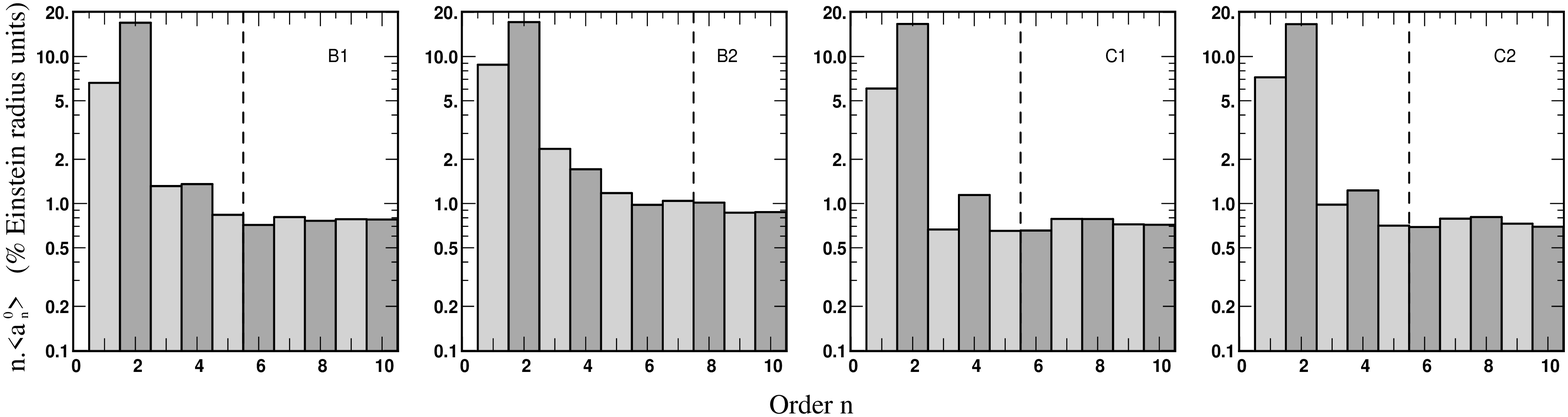}}
\caption{Variation of mean amplitudes $n\times \langle a^0_n \rangle $ for each lenses
 catalogue. The dashed line represents the confident limit.}
\label{figff0}
\end{figure*}

Having estimated the statistical contribution of substructures to the
perturbative fields, and we may now  address the problem
of defining observational signatures of substructures. The observational
effects are of two types: (i) an effect on the position of images
which is controlled by the field  \fone, and (ii) effects on the image
morphology, for instance its size in the orthoradial direction which
depends on the structure of the field \fzero. An
interesting point is that effects of type (i) are directly related
to \fone, while type (ii) effets is not related directly to \fzero
 but to its derivative (at least for images of  small extension). 
Indeed, let us 
consider a source  with circular contour; provided the image is
small enough, a local linear expansion of the field \fzero
 will be sufficient to estimate the image morphology. We make
the following field model:
\begin{equation}
  \frac{d\widetilde{f_0}}{{\rm d}\theta} \sim
 \frac{{\rm d}^2\widetilde{f_0}}{{\rm d}\theta^2} (\theta - \theta_0)\,,
\end{equation}
where $\theta_0$ is a position angle which should be close to the image center.
At the edge  of the images, we have:
\begin{equation}
  R_0 = \frac{d\widetilde{f_0}}{{\rm d}\theta}\,.
\end{equation} 
By defining $\Delta \theta= \theta - \theta_0$, it follows that:
\begin{equation}
  \frac{{\rm d}^2\widetilde{f_0}}{{\rm d}\theta^2} \Delta \theta = R_0\,.
\end{equation}
Thus, $({{\rm d}^2\widetilde{f_0}}(\theta)/{{\rm d}\theta^2})^{-1}$ scales
like the orthoradial image size  $\Delta \theta$
and can be directly related to observational quantities. The main
difference with effects of types (i), which are related to the field
rather than the field derivative, is that the derivative introduce
heavy weights on the higher orders of the Fourier serie expansion of
the field, where the substructure contribution is dominant. Indeed,
the derivative of the Fourier serie introduce a factor $n$ at order
$n$, which translates in a factor $n^2$ on the components of the power
spectrum. Thus  the image morphology (and in
particular its orthoradial extension for smaller images) will be much more
sensitive to substructure than the average image position.
To illustrate this, we have plotted in Figure (\ref{figff0}) the
amplitudes  $\langle a^0_n \rangle$ for lenses samples $B1$, $B2$, $C1$ and $C2$.  
Moreover, 
to study the contribution of order $n \geq 3$, it is convenient to define
the following quantities:
\begin{eqnarray}
  {\cal P}_1 &=& \langle (a^0_1)^2 \rangle  + 4\times \langle (a^0_2)^2 \rangle   \cr \cr 
  {\cal P}_2 &=& \sum_{n\ge 3} n^2\times \langle (a^0_n)^2 \rangle \cr \cr
  Q &=& \sqrt{\frac{{\cal P}_2}{{\cal P}_1}} \,,
  \label{}
\end{eqnarray}
%
Table~\ref{tab:Q} summarize mean values of $Q$  relative to each
catalogue of lenses. As expected, $Q$ has higher values when
substructure have a cusp profile.  The total contribution of high order
vary between $\sim 8$ and $18 \%$ according the model of lenses used,
which is quite significant.
\begin{table}
\begin{center}
\begin{tabular}{c|c|c|c|c|c}
\hline
 Sample & $k_0$ & $\alpha_0$ & $k_1$ & $\alpha_1$ & $Q$\\
\hline
$A$ & -  & -  & - & - & - \\
\hline
$B1$  &3.57 &-1.84  & 2.21 &-1.69 &0.113 $\pm$ 0.057   \\
\hline
$B2$  &7.33 &-2.07 &6.65 &-2.21  &0.179 $\pm$ 0.097 \\
\hline
$C1$  &0.74 &-0.950 & 0.546&-0.845 &0.083 $\pm$ 0.040  \\
\hline
$C2$  &2.08 & -1.575&1.462 & -1.468&0.095 $\pm$ 0.049 \\
\hline
\end{tabular}
\end{center}
\caption{Fit parameter of the  statistical distribution of the harmonics of the
two fields. \label{tab:Q}}
\end{table}
 In closing, for the sample $C2$, mean errors between predicted
 contours from the perturbative method and ray-tracing solution are  
 ${\rm err}= (0.95\pm 0.47) \% R_E$,
 ${\rm err}= (0.97\pm 0.48) \% R_E$ and
 ${\rm err}= (1.11\pm 0.49) \% R_E$  for isophotes
$0.6$, $0.2$ and $0.01 I_{\rm max}$ respectively.

\section{Conclusions}

The structuration of matter on galactic scales remains a privileged
framework to test cosmological models. In particular, the large number
of dark matter subhalos predicted by the $\Lambda$CDM cosmology is still
a matter of debate.
In this paper, potential signatures of substructures in the strong
lensing regime were considered. This investigation makes use of  the perturbative
solution presented by Alard (2007, 2008), in which small deviations from the
``perfect ring'' configuration are treated as perturbations.  
In this
framework, all informations on the deflection potential are contained in
two one-dimensionnal fields, $f_1(\theta)$ and ${{\rm d}f_0(\theta)}/{{\rm d}\theta}$
which are related to the radial expansion of the perturbed aspherical potential. The
analysis of the properties of these two fields via their harmonic decomposition represents a simple and
efficient way to track back possible signatures of subtructures.  
For instance,  the perturbative method offer a simple
and clear explanation of the multiplicity of the images.
As shown in Alard (2008), the unperturbed image
is a long arc for a cusp caustic configuration, but when a substructure
is introduced near the arc, this latter is broken in 3 images. This effect is also
illustrated in the present work in the example shown in the
figure (\ref{fig_s2}). Basically the breaking of
the image and the change in image mutiplicity is due to the perturbative
terms introduced by the substructure on the field ${{\rm d}f_0(\theta)}/{{\rm d}\theta}$.
In the case of a circular source with radius $R_0$, the image is broken when
${{\rm d}f_0(\theta)}/{{\rm d}\theta} > R_0$.

In this paper, lenses were modelled either by dark matter halos extracted from
cosmological simulations or via toy models. The advantage of toy
models is to reach a higher resolution and to allow us to study the influence of free parameters
such as the inner profiles of substructures which are expected to play a
central role here.
We have first estimated the accuracy of the perturbative predictions
by comparing the mean error between predicted contours images 
and theoretical ones derived via ray-tracing. We found that
in general, the relative mean error is $\sim 1\%$ of the Einstein
radius ($R_E$) when different impact parameters configurations and
different source redshifts are considered.
We also found  that both resolution limitation and
ray-tracing procedure lead to a systematic error of  $\sim 0.3\% R_E$. 
Furthermore, although the accuracy of this approach for elliptical lenses
  is demonstrated in Alard (2007), we have checked and shown in the present study
that the method works for the
range of ellipticities derived from realistic 3D dark matter halos.
This implies that the numerical evaluation of the coefficient of the 
perturbed potential,  \fone and
\fzero at the Einstein radius is accurate enough 
to carry a statistical investigation. 

We have generated several mock catalogues of  lenses  in which all objects have a
total mass of $10^{14} M_\odot$ and are at redshift
$z=0.2$. This value is motivated by comparison with
observational surveys, and is close to where the strong lensing
efficiency of clusters is the largest  for sources $z_s \geq 1$ (Li et
al. 2005).
The first catalogue represents our 
reference sample since all lenses are modelled by dark matter halos
without substructures. In the other ones, substructures are described
by either a cusp profile or a core
profile. Their radial distribution is also a free parameter and we have
used $C_{\rm sub} = C_{\rm host}$ and $C_{\rm sub} = 5$.
Our statistical investigation involves a  Monte-Carlo draw: the ellipticity of
host halos, position of substructures, sources redshift for instance are
randomly derived according to specific distributions.  
We found that the harmonic power spectra of  \fone and
\fzero tend to develop a tail towards  the large harmonics
when substructure are  accounted for. This effect is more pronounced
when substructures have a cusp profile.

Several improvements of the present investigation are envisioned since the ultimate
goal of the method is to provide a clear estimate of the amount of substructures in
observations.
\begin{itemize}
\item Statistically, the properties of the cross-power spectrum
   of \fzero~and \fone~will be instructive.
  A clear characterisation of the covariance of these fields observed in Fig. \ref{fig_champs}
  along with the result of Alard (2008) upon which, as high multipole order $n$, about the
  same power is contained in each of the fields, would allow us to reduce the dimensionality
  of the problem and perhaps only consider either \fone~or \fzero.
\item In this vein, it remains to be shown to which extent the statistical properties of
  \fone~and \fzero~can be approximated as Gaussian random fields. If so, the realization of
  mock giant arcs would be greatly simplified. Concerning this point, 
      the number of experiments of the present work must be increased to provide a clear diagnostic. 
\item A natural extension of this work would be to consider more realistic
  lenses while taking into account the dynamics of substructures inside the
  host halo, its connection to 
  cosmology via the expected statistical distribution  of substructures
  (see e.g. Pichon \& Aubert 2006),  as well as star formation mechanisms.
  Indeed, stripping proccesses caused by tidal forces may lead to more complex structures.
  For instance, the study of some merger events in the phase-space (radial
  velocity versus radial distance) reveals the formation of structures
  quite similar to caustics generated in secondary infall models of halo
  formation (Peirani \& de Freitas Pacheco 2007).
  On the other hand, cooling processes and subsequent star formation may
  lead to steeper dark matter profile due to adiabatic contraction. 
  Then, such processes should have an impact in the amplitude of high
  orders of our study.
\item Obviously a perturbation of the simple circularly symmetrical case needs not lie in
  the same lens plane as the main deflecting halo. Uncorrelated halos superposed along the
  line of sight to the background source may well introduce perturbations on top of
  substructures belonging to the main halo. This will contribute to some additional
  shot noise background in the power spectra of \fone~and \fzero, which has to be quantified
  and subtracted off using ray-tracing through large simulated volumes. This work is beyond
  the scope of the present analysis.
\item On the path to a possible inversion yielding \fone~and \fzero~from observed
  arcs shapes and locations, several unknowns left on the rhs of Eq. \eqref{eq_invert_0} have
  to be controlled and will need to be fitted for in a non-linear way in order to attempt
  a reconstruction of fields \fone~and \fzero.
  In addition, we only have considered a simple representation of the background source.
  A lumpier background source will translate into a less regular arc with some small scale
  signature in the observable quantities such as $\der r_\pm$.
 However, the replication of these internal fluctuations along the arcs and,
  possibly, in the counter image, as well as the
  information contained in the various isophotes
  could allow us to
  reconstruct \fone and \fzero directly from the observations.
 In this respect, Alard (2008) provides a general inversion method 
 when two circular sources are considered for instance.
\end{itemize}

To conclude, the upcoming generation of high spatial resolution
instruments dedicated to cosmology (\eg JWST, DUNE, SNAP, ALMA)
will provide us with an unprecedented number of giant arcs at all scales.
The large samples expected will make standard lens modellings untractable and require
the development of new methods able to capture the most relevant source of constraints
for cosmology. In this respect, the perturbative method we present here may turn out to be
a promising research line.


\section{Acknowledgements}

S. P. acknowledges the financial support through a ANR grant.
This work made use of the resources available within the framework of
the horizon collaboration: \texttt{http://www.projet-horizon.fr}. 
It is a pleasure to thanks T. Sousbie, K. Benabed, S. Colombi, B. Fort
and G. Lavaux for interesting conversations.
 We thank
 the referee for his useful comments that helped to improve
 the text of this paper.
We would  also like to thank D.~Munro for
freely distributing his Yorick programming language (available at
\texttt{http://yorick.sourceforge.net/}) which was used during the
course of this work.




\label{lastpage}

\end{document}